# Electrical spin injection into InGaAs/GaAs quantum wells: a comparison between MgO tunnel barriers grown by sputtering and molecular beam epitaxy methods

P. Barate[1], S. Liang[2], T. T. Zhang[1], J. Frougier[3], M. Vidal[1], P. Renucci[1], X. Devaux[2], B. Xu[4], H. Jaffrès[3], J. M. George[3], X. Marie[1], M. Hehn[2], S. Mangin[2], Y. Zheng[5], T. Amand[1], B. Tao[2,6], X. F. Han[6], Z. Wang[4], Y. Lu[2*]

[1]Université de Toulouse, INSA-CNRS-UPS, LPCNO, 135 avenue de Rangueil, 31077 Toulouse, France

[2]Institut Jean Lamour, UMR 7198, CNRS-Nancy Université, BP 239, 54506 Vandoeuvre, France

[3]Unité Mixte de Physique CNRS/Thales and Université Paris-Sud 11, 1 avenue A. Fresnel, 91767 Palaiseau, France

[4]Key Laboratory of Semiconductor Materials Science, Institute of Semiconductors, Chinese Academy of Sciences, P. O. Box 912, Beijing 100083, China

[5]Institut des NanoSciences de Paris, UPMC, CNRS UMR 7588, 4 place Jussieu, 75005 Paris, France

[6]Beijing National Laboratory for Condensed Matter Physics, Institute of Physics, Chinese Academy of Sciences, P.O. Box 603, Beijing 100190, China

*Corresponding author: yuan.lu@univ-lorraine.fr

**Abstract**

An efficient electrical spin injection into an InGaAs/GaAs quantum well light emitting diode is demonstrated thanks to a CoFeB/MgO spin injector. The textured MgO tunnel barrier is fabricated by two different techniques: sputtering and molecular beam epitaxy (MBE). The maximal spin injection efficiency is comparable for both methods. Additionally, the effect of annealing is also investigated for the two types of samples. Both samples show the same trend: an increase of the electroluminescence circular polarization ($P_c$) with the increase of annealing temperature, followed by a saturation of $P_c$ beyond 350°C annealing. Since the increase of $P_c$ starts well below the crystallization temperature of the full CoFeB bulk layer, this trend could be mainly due to an improvement of chemical structure at the top CoFeB/MgO interface. This study reveals that the control of CoFeB/MgO interface is essential important for an optimal spin injection into semiconductor.

Efficient injection of spin-polarized electrons from a ferromagnetic (FM) source into a semiconducting heterostructure is a prerequisite for the realization of a large number of new spintronic devices including spin-transistors[1], spin light emitting diodes (spin-LEDs)[2-12] and spin-lasers[13]. It has been shown that inserting a thin MgO layer between the FM contact and semiconductor can circumvent the problem of conductivity mismatch between FM metals and semiconductors, resulting in a very high spin injection efficiency[14]. In particular, the CoFe/MgO injector has exhibited the highest spin injection yield at room temperature up to now (the electroluminescence polarization can reach 32% at 300K)[6]. This magnetic tunnel injector is constructed with two interfaces: the ferromagnetic metal (FM)/MgO and the MgO/semiconductor interfaces. It is still an open question to determine the role of interface for an optimal spin injection. Moreover, although the symmetry rules for the tunneling transfer of carriers in magnetic tunnel junctions (MTJs) with an MgO tunneling barrier[15-17] are well known for standard MTJ systems, it is significantly less understood if the MgO can act as a relevant spin filter to enhance the spin injection efficiency into semiconductor. In this letter, we propose to tackle these points by studying spin-LED systems with CoFeB/MgO/GaAs injectors, where the MgO barrier is grown by two different techniques: sputtering and molecular beam epitaxy (MBE). This results in the different quality of MgO/GaAs and CoFeB/MgO interface prepared by the two different techniques, which can allow us to examine which interface is crucial for spin injection. In addition, CoFeB is a promising candidate as a ferromagnetic injector, because after 350°C annealing CoFeB is easy to form a bcc grain-to-grain epitaxial crystalline phase on MgO18. This improvement by annealing has been particularly observed in MTJs, where 600% tunnel magnetoresistance (TMR) was measured on CoFeB/MgO/CoFeB junctions[19] instead of 180% observed on CoFe/MgO/CoFe junctions[17]. Studying the annealing effect on these spin-LED with CoFeB/MgO injectors by correlating the circular light polarization with the crystallization state of CoFeB[20,21] will allow us to examine the key role of interface and further determine if the MgO spin filtering effect exists for spin-injection into semiconductors.

The quantum well (QW) *p-i-n* LED device has the following structure sequence (Fig. 1): *p*-GaAs:Zn (001) substrate ($p=2\times10^{19}$cm$^{-3}$) /500 nm *p*-GaAs:Be ($p=2\times10^{19}$cm$^{-3}$) /200 nm *p*-GaAs:Be ($p=2\times10^{18}$cm$^{-3}$) /50 nm undoped GaAs /10 nm undoped In$_{0.1}$Ga$_{0.9}$As /50 nm undoped GaAs /50 nm *n*-GaAs:Si ($n=1\times10^{16}$cm$^{-3}$). The LED surface was passivated with arsenic in the III-V MBE chamber and then transferred through air into another MBE-sputtering interconnected system. The arsenic capping layer was firstly desorbed at 300°C in the MBE chamber. Two methods were then used to grow the MgO tunneling barrier layer. Either MgO is grown at 250°C in the MBE chamber after arsenic desorption, or the sample is transferred into the sputtering chamber through vacuum to grow the MgO layer. In both cases, MgO layer has identical thickness of 2.5nm. Finally, the 3nm CoFeB ferromagnetic contact and 5nm

Ta protection layer were deposited by sputtering. Hereafter, we will name the samples "MBE" and "sputtering" to refer to two different spin-LEDs with MgO prepared by MBE and sputtering techniques, respectively. These two growth techniques produce a different quality of MgO barrier and also a different quality of interface (see in Fig. 1 "top and bottom interface"). It will allow us to determine which factor is crucial for the efficient electrical spin injection, knowing that the thicknesses of all layers involved in the injector are the same for both types of sample. 300 μm diameter circular mesas were then processed using standard UV photolithography and etching techniques. In the end, the processed wafers were cut into small pieces to perform rapid temperature annealing (RTA) at different temperatures ($T_a$) for one minute. The RTA procedure is a good way to promote the crystallization of CoFeB[22] while preventing a change to the LED optical characteristics.

The electroluminescence (EL) results are a first test of the difference between the two barrier growth methods. A more detailed work condition of the polarization resolved EL measurements can be found in a previous work[8] by our group. Insets of Fig. 2(a) show a typical CW EL spectra from a spin-LED with a sputtered MgO tunnel barrier ($T_a$=350°C) acquired at 25K under a bias of $V_{bias}$=2.4V for $B$=0T (top inset) and $B$=0.8T (bottom inset). The EL circular polarization $P_c$ is defined as $P_c=(I^{\sigma+}-I^{\sigma-})/(I^{\sigma+}+I^{\sigma-})$ where $I^{\sigma+}$ and $I^{\sigma-}$ are the intensities of the right and left circularly polarized components of the luminescence, respectively. Whereas the heavy-hole exciton (XH) EL peak observed at 878nm does not show any circular polarization at zero magnetic field (top inset), the $P_c$ reaches 24% ± 1% under $B$=0.8T (bottom inset). Additionally, the magnetic circular dichroism is less than 1% at 0.8T thanks to a measurement based on a linearly polarized He-Ne laser[23]. The measured $P_c$ increases with the applied longitudinal magnetic field [Fig. 2(a)], due to the progressive increase of the projection of the out-of-plane magnetization (the magnetization of CoFeB layer is within the plane at zero magnetic field), which can be explained by the optical selection rules applied to the QW[24]. In Fig. 2(b), we present a systematic study of the influence of the post-annealing temperature on $P_c$ for the two kinds of spin-LED samples. For sputtered tunnel barriers, we observe a clear improvement of the measured EL circular polarization rate from 13.5±1% before annealing up to 24±1% for the optimal $T_a$=350°C, following by a decrease of $P_c$ with $T_a$=380°C. This trend is similar for the sample based on a MBE grown barrier, with a slightly lower optimal polarization of 20±1% [the corresponding spectra are shown in the inset of Fig. 2(b)]. The behavior with annealing is similar to the one observed by Wang *et al.*[25] for spin-LED with CoFe/MgO injectors, but the variation in our spin-LED is much remarkable.

The use of spin-LEDs as an optical means to quantify the electrical spin injection in GaAs is based on a straightforward relationship between the electron spin polarization $P_e$ injected in the QW and the measured EL circular polarization $P_c$: $P_c=P_e*F$ [26,27]. The $F$ factor takes into account the electron spin relaxation in the QW during the electron lifetime: $F=1/(1+\tau/\tau_s)$ with $\tau_s$ spin relaxation time and $\tau$ electron lifetime in the QW[26,27]. It is important to determine if the observed large improvement of $P_c$ as a function of annealing temperature displayed in Fig. 2(b) is due to a real improvement of $P_e$, or only due to the impact of the annealing process on the QW properties. Therefore, a systematic measurement of the $F$ factor as a function of the annealing temperature was performed by measuring $\tau_s$ and $\tau$ in a bare *p-i-n* LED sample by time and polarization resolved photoluminescence (TRPL). Detailed experimental conditions can be found elsewhere[8]. As shown in Fig. 2(c) as an example for the bare *p-i-n* LED with $T_a=350°C$, the electron lifetime $\tau$ can be determined from the decay of the sum of $I^{\sigma+}$ and $I^{\sigma-}$ to be 180±10 ps at 1/e, and the spin relaxation time $\tau_s$ can be determined from the decay of the circular polarization to be 425±50 ps at 1/e. So the $F$ factor can be determined to be 0.7. The measured $\tau$ and $\tau_s$ and the deduced $F$ factor as a function of $T_a$ are summarized in the inset of Fig. 2(d). For better comparison of the variation of $P_c$ and $F$ factor as a function of $T_a$, the relative change compared to the case before annealing (BA) is plotted in Fig. 2(d). It clearly shows that the relative improvement of the circular polarization $[P_c(T_a)-P_c(BA)]/P_c(BA)$ reaches about 80% between $P_c(BA)$ and $P_c(T_a=350°C)$ for sputtered samples. However, the relative variation of the $F$ factor $[F(T_a)-F(BA)]/F(BA)$ is much weaker [less than 10% between $F(BA)$ and $F(T_a=350°C)$] and is even negative. So the large improvement of $P_c$ observed in Fig. 2(b) is due to a real improvement from the spin injector part ($P_e$) because of the small (and opposite) variation of the $F$ factor.

In the EL measurement, the maximum out-of-plane field is limited at 0.8T, which cannot fully saturate the CoFeB magnetization. To eliminate the possibility of the increase of $P_c$ due to the change of the saturation field after annealing, a superconducting quantum interference device magnetometer (SQUID) was employed to measure the magnetization of film sample of a sputtered spin-LED in out-of-plane configuration. As shown in Fig. 3(a), the magnetization is quasi linear between -1.3T and 1.3T as a function of magnetic field for before annealing and $T_a=275°C$ and becomes non-linear with $T_a=350°C$. Let's note that the saturation field measured is ~1.3T for $T_a$ below 300°C and ~1.75T for 350°C. In order to extrapolate $P_c$ at saturation, the $P_c$ value at 0.8T is multiplied by a ratio $M_{sat}/M(0.8T)$ based on the results obtained by SQUID. The extrapolation leads to a rough estimation of $P_c$ at saturation of about 42.0% at 25K for spin-LED with sputtered MgO after 350°C annealing, proving the high efficiency of CoFeB/MgO injector. This result is close to the best obtained by Jiang *et al.* ($P_c=50\%$ at low temperature with CoFe/MgO/AlGaAs injector[6]) and is larger than the one reported very recently by Li *et al.*[28] with a Schottky barrier as spin injector ($P_c=25\%$). As shown in Fig. 3(a) with dashed line, the ratio $M_{sat}/M(0.8T)$ does not depend on the annealing temperature, which validates the direct comparison of $P_c$ measured at 0.8T as function of $T_a$ in Fig. 2(b). The non-linear *M-H* curve for $T_a$=350°C in fact reflects the crystallization of CoFeB layer[20]. To further clarify the influence of the annealing temperature on the crystallization of CoFeB layer, SQUID measurements were performed in an in-plane configuration to check the coercivity $H_c$ of CoFeB layer. Fig. 3(c) displays the hysteresis cycles of two types of spin-LEDs annealed at different temperatures. Fig. 3(b) summarizes the evolution of $H_c$ as a function of $T_a$. When $T_a$ is above 300°C, $H_c$ increases in both types of samples. This is a strong indication of the beginning of the CoFeB crystallization[22]. It is noted that the MBE sample has a relative larger $H_c$ than the sputtering sample after annealing, which could be related to the different magnetic domain structures in CoFeB layer (discussed below). The increase of the coercivity field with $T_a$ saturates around 350°C, signifying a full crystallization of CoFeB layer, which is in good agreement with reported crystallization temperature[20]. However, these EL measurements indicate that $P_c$ increases begin far below 300°C, with saturation at 350°C when the crystallization is finished. Therefore it is clear that the increase of $P_c$ is not due to the crystallization of the whole CoFeB layer.

To better understand the effect of annealing on $P_c$, high-resolved transmission electron microscopy (HR-TEM) is used to examine the interfacial structural information. The good homogeneity of the structures is checked on the low magnification images in the insets of Fig. 4(d) [for the sputtering sample annealed at 350°C, the MBE sample has identical feature (not shown)]. Note also that the MgO thicknesses are identical for the two types of samples validating the direct comparison of the measured circular polarization from the two kinds of spin-LEDs in Fig. 2(b). First, we compare the TEM images for MBE samples before annealing [Fig. 4(a)] and after annealing at 350°C [Fig. 4(b)]. In both cases the MgO/GaAs interface is sharp. The CoFeB layer is amorphous before annealing and fully crystallized after 350°C annealing, which further confirms the increase of $H_c$ measured by SQUID is due to the crystallization of CoFeB layer. The selected zone FFT image shows a bcc structure of CoFe(B) on MgO [left inset of Fig. 4(b)]. The right inset of Fig. 4(b) shows the bright field HR-scanning TEM image, which allows us clearly identify the epitaxial crystalline orientation relationship between GaAs, MgO and CoFe (B is absorbed by Ta): GaAs[100](100)//MgO[100](100)//CoFe[110](010). As $P_c$ clearly increases with $T_a$ for the MBE sample [Fig. 2(b)] whereas no observed change is found at the MgO/GaAs interface and while a drastic change occurs at CoFeB/MgO interface (evidenced by the CoFeB crystallization at 350°C), one can infer that the CoFeB/MgO interface is the crucial one for an optimal spin injection. A second observation reinforces this conclusion: in a comparison of the TEM images for the MBE [Fig. 4(a)] and sputtering [Fig. 4(c)] samples before annealing, a presence of a thin amorphous layer (~0.4 nm) is detected at the MgO/GaAs interface (bottom interface) for the sputtering sample [Fig. 4(c)]. The fact that $P_c$ is comparable [Fig. 2(b)] before annealing for both samples (and even slightly better for a sputtered MgO barrier) whereas the quality of the MgO/GaAs interface is much better by MBE also indicates that the influence of the MgO/GaAs interface is weak and that the CoFeB/MgO interface is the most important one for an efficient electrical spin injection. Finally, for the sputtering samples after 350°C annealing [Fig. 4(d)], the fully crystallization of CoFeB layer is also confirmed by the FFT image [left inset of Fig. 4(d)] as same as the MBE sample. It is also noted that the amorphous layer at MgO/GaAs interface is much reduced after annealing. This amorphous layer could be due to the large kinetic energy of bombarded atoms during the sputtering growth process, which is then recrystallized after annealing.

To complete this study, reflection high energy electron diffraction (RHEED) measurements have been performed to check the MgO surface crystalline structure (top interface) prepared by the different methods. For MBE sample [insets of Fig. 4(a)], the RHEED patterns along GaAs [100] and [110] directions exhibit a monocrystalline spotty diffraction pattern, which further confirms the same in-plane epitaxial relationship (GaAs[100](100)//MgO[100](100)) as that obtained from the HR-STEM image. However, for the sputtered MgO surface the RHEED images [insets of Fig. 4(c)] indicate a worse crystalline quality since some polycrystalline rings appear in the two directions. Nevertheless, the fact that the values of $P_c$ are slightly larger for sputtering samples (whatever the annealing temperature is) despite this worse MgO quality indicates that the texture quality of the MgO barrier is less crucial for spin injection. As well known, during annealing, the Ta layer can absorb B atoms which results in the crystallization of CoFeB from the MgO/CoFeB interface[29]. For the sputtering sample, the polycrystalline MgO grains observed before annealing will certainly induce high density of grains in the crystallized CoFeB layer after annealing due to the grain to grain epitaxial procedure[18]. The grain boundary related magnetic domain structures[30] could explain the smaller coercivity observed in SQUID measurements [Fig. 3(b)] for the sputtering sample compared to the MBE sample.

To conclude, the quality of the CoFeB/MgO interface seems to be crucial for an efficient electrical spin injection. As a large increase of the spin injection efficiency takes place below annealing at 300°C, *i.e.* before crystallization of the whole CoFeB layer, we attribute this trend to an improvement of chemical bounds at the CoFeB/MgO interface, as it was also observed for the TMR improvement in MgO MTJs below 300°C annealing[31]. The sub 300°C annealing can move the Co and Fe atoms toward to the O atom at CoFeB/MgO interface, which efficiently enhances the interfacial spin polarization and interfacial perpendicular magnetization anisotropy[32]. Another point: whatever the annealing temperature is, the values of $P_c$ are slightly larger for sputtering samples which undergo no interruption during the growth of the CoFeB/MgO interface (while the MBE samples require a pause for sample transfer). This is another proof of the high importance of this interface. From 300°C to 350°C annealing, since the improvement of $P_c$ is marginal, the MgO spin filter effect which can select the symmetry $\Delta_1$ electrons in CoFe band structure (possessing highest spin-polarization for tunneling) does not play an important role in our case. This could be related to the Ta diffusion toward CoFeB/MgO interface[19,33], which happens after high temperature annealing as indicating from the decrease of $P_c$ with $T_a$=380°C [Fig. 2(b)]. Therefore, future enhancement of the MgO spin filtering effect, may rely upon the suppression the Ta diffusion (*eg.* replace with other metals for B absorption).

In summary, an efficient electrical spin injection from CoFeB into InGaAs/GaAs QW through MgO tunnel barriers is demonstrated ($P_c$~24±1% at 0.8T corresponding to $P_c$~42% at saturation at 25K). Spin-LEDs with MgO tunnel barriers are fabricated by two different techniques (sputtering and MBE). A systematic study of the impact of the post-annealing temperature on the two types of samples shows a similar increase of $P_c$ with the increase of the annealing temperature and a comparable optimized spin injection efficiency for both methods in the range of 300-350°C. This behavior is mainly attributed to the improvement of chemical structure at the top CoFeB/MgO interface because the increase of $P_c$ starts far below the crystallization temperature of the whole CoFeB layer.

**Acknowledgements**

We thank T. Hauet for the help of SQUID measurement. We also acknowledge Prof. Mair Chshiev for helpful discussion on MgO spin filtering effect. This work was supported by the France-China ANR-NSFC research project SISTER (Grant No. ANR-11-IS10-0001), and French ANR research project INSPIRE (Grant No. ANR-10-BLAN-1014). X. F. Han also acknowledges the support of Chinese State Key Project of Fundamental Research of Ministry of Science and Technology (MOST, No. 2010CB934401).

**Figure captions**

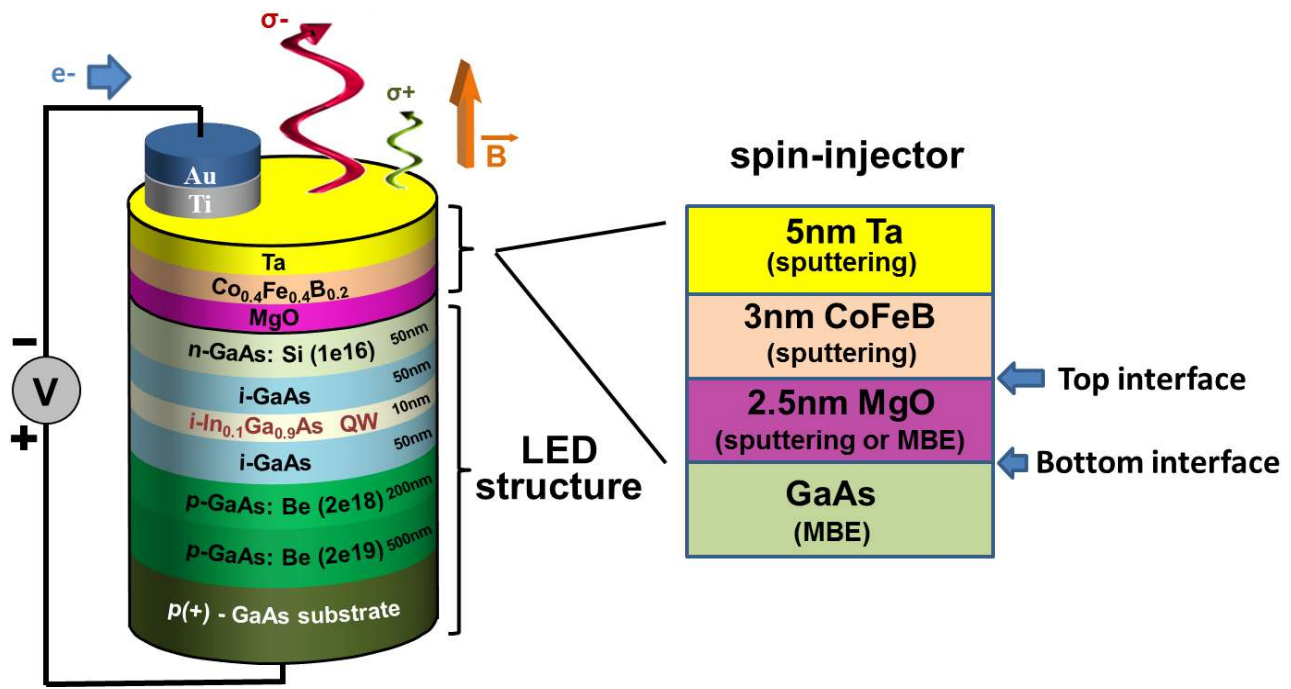

FIG. 1. (Color online). Spin-LED structure with a single InGaAs/GaAs QW. The CoFeB/MgO/GaAs spin injector layers are decomposed in terms of a top interface CoFeB/MgO and a bottom interface MgO/GaAs.

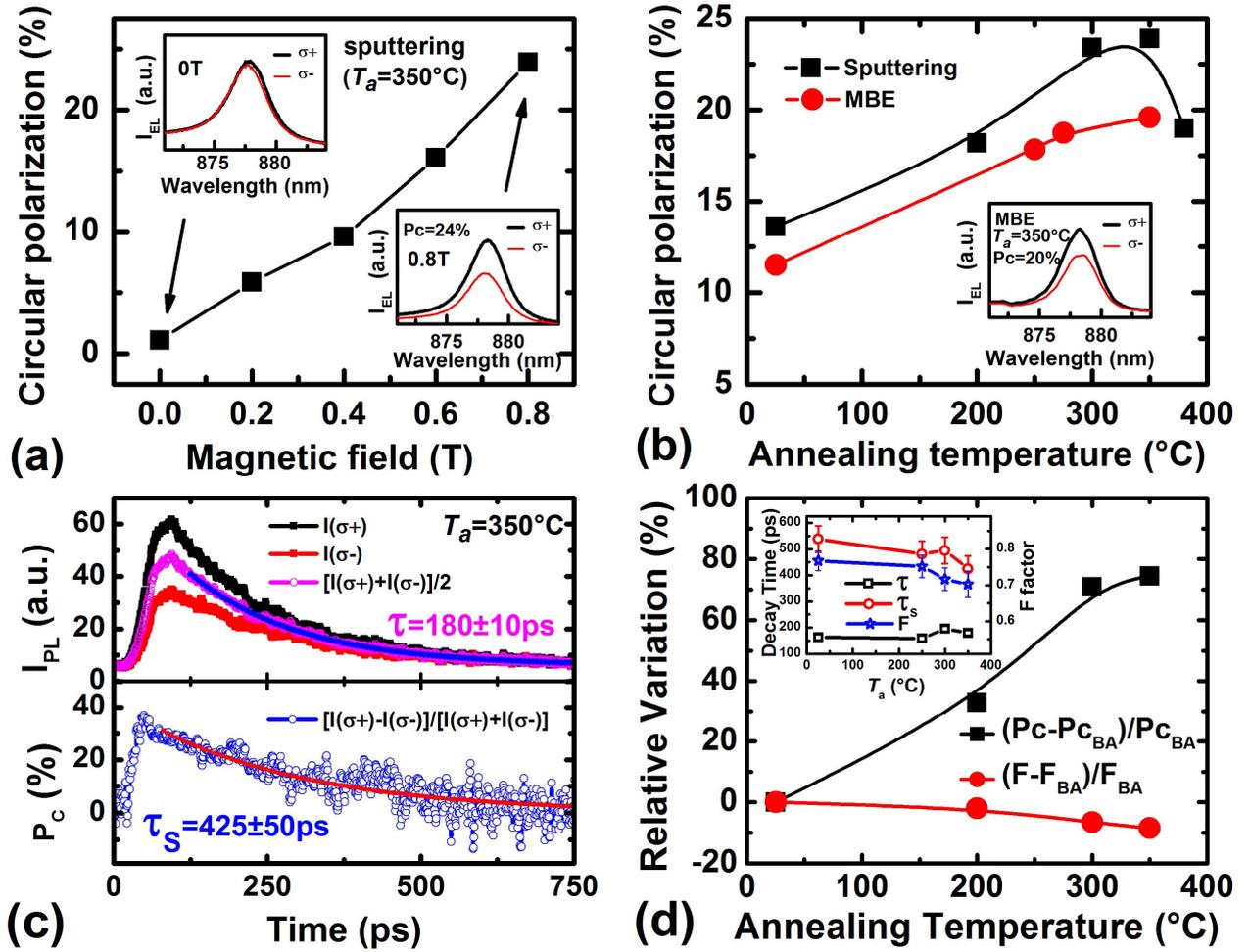

FIG. 2. (Color online) (a) EL circular polarization $P_c$ as a function of the applied longitudinal magnetic field for the spin-LED with a sputtered MgO tunnel barrier ($T_a$=350°C) at $T$=25K. EL spectra at zero magnetic field (top inset) and at $B$=0.8T (bottom inset) for $I^{\sigma+}$ (thick black line) and $I^{\sigma-}$ (thin red line) EL components. (b) EL circular polarization $P_c$ as a function of the annealing temperature for sputtering (black squares) and MBE (red circles) grown MgO spin-LEDs at $T$=25K. Inset: EL spectra of a spin-LED with a MBE grown tunnel barrier ($T_a$=350°C) at $T$=25K and $B$=0.8T for $I^{\sigma+}$ (thick black line) and $I^{\sigma-}$ (thin red line) EL components. (c) TRPL measurement on a bare p-i-n LED sample ($T_a$=350°C) at $T$=25K. Top: PL intensity components $I^{\sigma+}$ and $I^{\sigma-}$ with respectively $\sigma^+$ polarization (black squares) and $\sigma^-$ polarization (red squares) as a function of time after a 1.5ps laser ($\sigma^+$) pulsed excitation at 780 nm (above GaAs bandgap). The sum of the two intensities components $I_{sum}= I^{\sigma+}+I^{\sigma-}$ is displayed in pink opened circles. The decay is characterized by the electron lifetime $\tau$. Bottom: time evolution of $P_c$ of PL at $T$=25K (blue opened circles). The decay is characterized by the electron spin relaxation time $\tau_s$. (d) Relative variation of the EL $P_c$ (black squares) for sputtered samples and of the $F$ factor (red circles) as a function of $T_a$. Inset: Electron spin relaxation time $\tau_s$ (red open circles), electron lifetime $\tau$ (black open squares) and $F$ factor (blue open stars) as a function of the annealing temperature $T_a$.

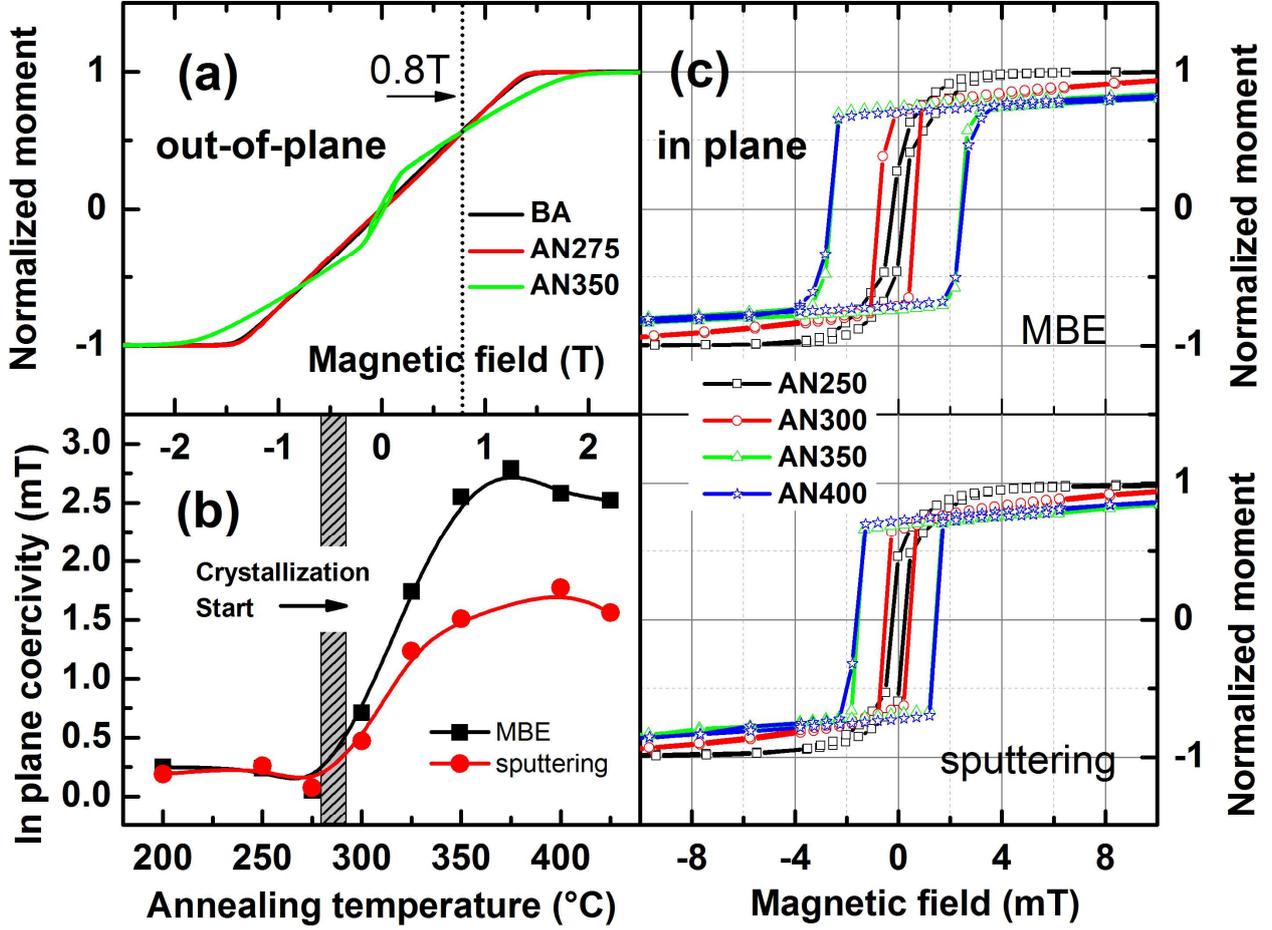

FIG. 3. (Color online) SQUID measurement at $T$=300K. (a) Normalized magnetization as a function of the applied out-of-plane magnetic field for a sputtering spin-LED before annealing (black line) and annealed at 275°C (red line) and 350°C (green line). (b) In-plane coercivity field $H_c$ as a function of the annealing temperature $T_a$ for a spin-LED with sputtered MgO (red circles) and with MBE grown MgO (black squares). (c) Top: In-plane hysteresis loop of spin-LED with MBE grown MgO annealed at different temperatures. Bottom: Same measurements for a spin-LED with sputtered MgO tunnel barrier.

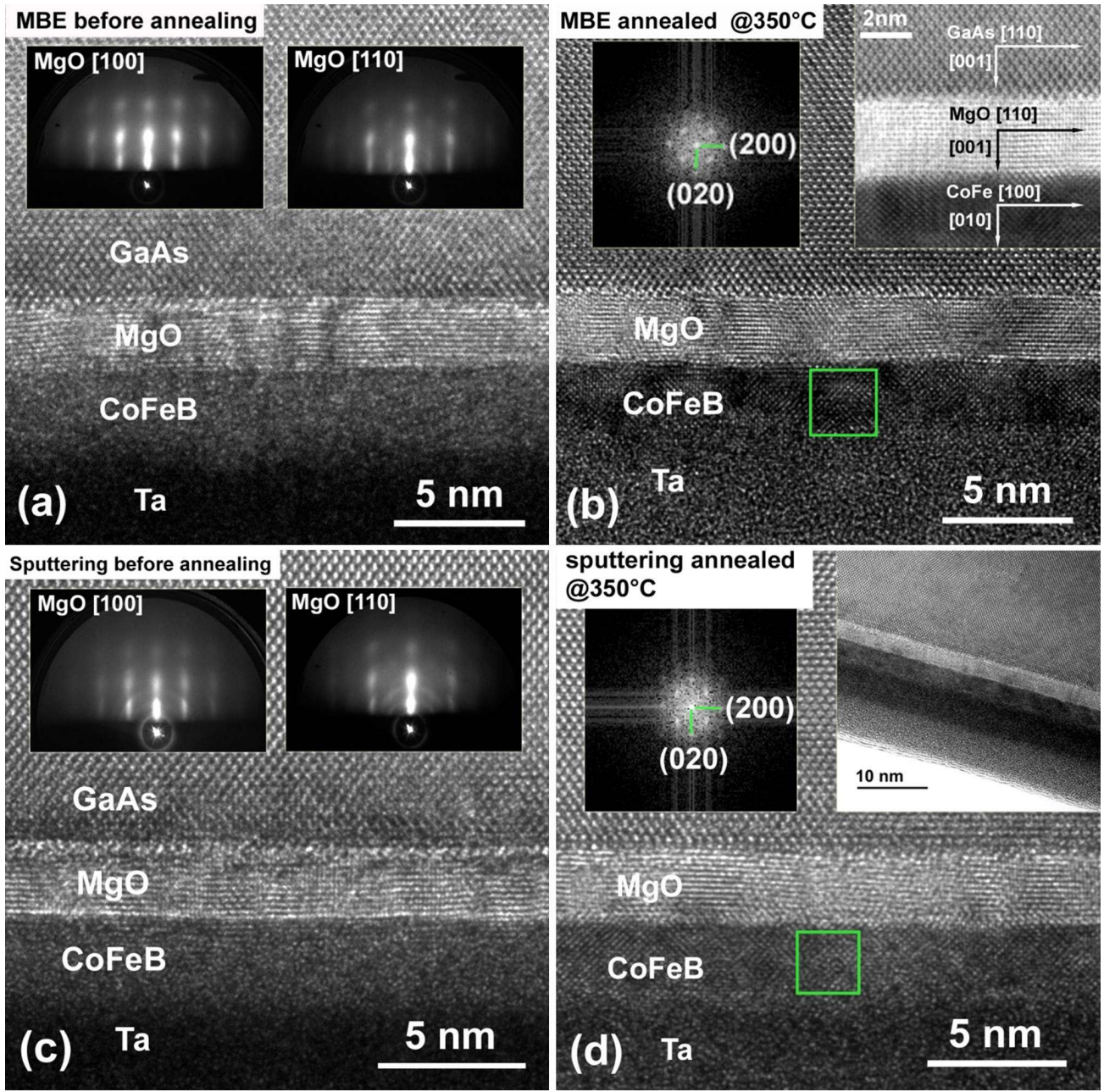

FIG. 4. (Color online) HR-TEM images of (a) sample with MBE grown MgO and (c) sample with sputtered MgO before annealing. Insets: RHEED patterns on MgO surface from GaAs [100] and [110] azimuths (left and right insets respectively). (b) HR-TEM image of the sample with MBE grown MgO after annealing at $T_a$ =350°C. Left inset: FFT pattern on the square zone in CoFeB layer. Right inset: HR-STEM image to show the crystallographic orientation relationship between GaAs, MgO and CoFe. (d) HR-TEM image of the sample with sputtered MgO after annealing at $T_a$ =350°C. Left inset: FFT pattern on the square zone in CoFeB layer. Right inset: TEM image with a large scale to show the homogeneity of structures.